\documentstyle[lat98,epsfig]{article}
\def\sec{$^{\prime\prime}$}
\begin{document}
\ \ 
\vskip 1.1cm

\title{NUCLEAR ACTIVITY IN NEARBY GALAXIES}
\author{Thaisa Storchi-Bergmann\altaffilmark{1}}
\altaffiltext{1}{Instituto de F\'\i sica, UFRGS, Brasil}

\begin{resumen}
I discuss some recent observational results in the research of nearby active galactic nuclei (AGN). These results cover three main topics: (i) evidences for the current paradigm for AGN's, which include a nuclear supermassive blackhole (SMBH) fed via an accretion disk; (ii) evidence that this paradigm may also apply to LINER's, the lowest luminous AGN's and to normal galaxies; (iii) evidences of how the fueling of the SMBH occurs and its relation to recent and intermediate age (10$^6$ to 10$^8$ yrs old) episodes of star formation.
\end{resumen}

\begin{abstract}
I discuss some recent observational results in the research of nearby active galactic nuclei (AGN). These results cover three main topics: (i) evidences for the current paradigm for AGN's, which include a nuclear supermassive blackhole (SMBH) fed via an accretion disk; (ii) evidence that this paradigm may also apply to LINER's, the lowest luminous AGN's and to normal galaxies; (iii) evidences of how the fueling of the SMBH occurs and its relation to recent and intermediate age (10$^6$ to 10$^8$ yrs old) episodes of star formation.
\end{abstract}

\keywords{\bf GALAXIES: ACTIVE --- GALAXIES: NUCLEI --- BLACKHOLES --- GALAXIES: STELLAR CONTENT}

\section{Introduction}

In the current paradigm (Lynden-Bell 1969; Blandford \& Rees 1992; Wilson 1996), active galaxies host a nuclear supermassive black hole fed via an accretion disk, which may also collimate a radio jet. The accretion disk is surrounded by the high density clouds of the Broad Line Region and a molecular torus
(Antonucci 1993), which would be responsible for hiding all these structures in the Seyfert 2 galaxies as well as for collimating the nuclear radiation incident on the Narrow Line Region.

The motivation for the above scenario begins with the large energies released by radio galaxies and Quasars ($\ge 10^{61}$ ergs) combined with the short time scale for the nuclear variability (down to minutes in X-rays), indicating a compact nuclear source with a very efficient mechanism for energy generation (more efficient than in stars, otherwise the whole stellar population of a 10$^{10}$ M$_\odot$ galaxy should explode as supernova in order to produce the necessary energy over the quasar lifetime). The efficiency for energy generation via accretion is estimated to be $\epsilon \sim 0.1$, as compared with 0.007 for nuclear burning in stars.

I find it interesting to make a parallel with what is observed in stars. 
Besides the inferred large masses for some X-ray sources in binary systems, additional evidences for a similar scenario as described above have become very convincing. Double-peaked emission lines from rotating emitting gas are clear signatures of the presence of an accretion disk (Kinney 1994 -- see the particularly convincing case of the eclipsing system observed by Marsh \& Horne 1990; Marsh 1999). More recently, Mirabel \& Rodriguez (1998) report the discovery of stellar ``microquasars'' in our galaxy, which present, on a smaller scale, many of the observed characteristics of quasars. In particular, recent simultaneous observations of the Gamma-ray source GRS1915+105 in X-rays, infrared and radio (Eikenberry et al. 1998) have shown periodic (P $\sim$30 min) X-ray flares which are followed by infrared and radio flares. 
The interpretation is that the X-ray flare is 
produced by the heated inner part of the accretion disk, followed by partial accretion and jet ejection, originating the IR and radio flares. These observations are, to date, the most clear evidence for the origin of the jets.

What are the evidences we have for the above scenario in AGN's?
There is no doubt about the presence of jets in radio galaxies and
Quasars. In this review, I concentrate on some recent evidences 
for the presence of nuclear SMBH's, accretion disks and tori.
As these are very small structures, the best evidences are found
in nearby galactic nuclei. Interestingly these nuclei comprise
not only ``classical'' AGN's, but also LINER's
and normal galaxies.

In the standard scenario, the nuclear luminosity is extracted from mass
accretion onto the hole. I also discuss observational evidences about
how the feeding may be occurring.

\section{Evidences for the standard scenario in AGN's}

\subsection{Torus}

Evidences for the presence of a dusty molecular torus in AGN's
are essentially those which led to the
Unified Scenario for Seyfert galaxies (Antonucci 1993), and include:
photon balance calculations (Kinney et al. 1991), showing that,
in Seyfert 2's the UV spectrum shows a deficit in the number
of ionizing photons when compared with that necessary to ionize the gas,
interpreted as due to blocking of the nuclear continuum by a torus seen
close to edge on; cone-like morphologies for the high excitation 
gas, interpreted as due to collimation of the ionizing radiation by
a nuclear dusty torus (Wilson 1992, 1997)l; Seyfert spectra seem to confirm
the predicted infrared and multiwavelength emission of the torus (Storchi-Bergmann, Mulchaey \& Wilson 1992; Mulchaey et al. 1994);
recent ASCA observations (Awaki 1997; Turner et al. 1997) have shown
that all Seyfert 2's exhibit equivalent hydrogen columns to their nuclei
($\sim$ 10$^{22.5}$ cm$^{-2}$) above those observed for Seyfert 1 nuclei. 

\subsection{SMBH}

The evidences for a nuclear SMBH have been obtained
essentially from stellar kinematics. The gaseous kinematics is
discussed bellow, in association with accretion disks.

Kormendy \& Richstone (1995), Rees (1998) and Ho (1998) present reviews about the evidences for the presence of a massive dark object (MDO) in the nuclei of galaxies from stellar kinematics. It is interesting to point out that in only
one ``classical'' AGN, M87, convincing evidence for the presence of a nuclear MDO using this method has been found. This result is  due to the fact that the influence of the SMBH extends to short distances from the hole, which can be estimated by $r\sim GM/\sigma^2$, where $M$ is the black hole mass and $\sigma$ is the velocity dispersion of the stars. In arcseconds, and using typical values for $M$ and $\sigma$ one obtains (Ho 1998)
$r\sim 1^{\prime\prime}(M/2\times 10^8 M_\odot)(D/5 Mpc)^{-1}(\sigma/200$ km s$^{-1})^{-2}$. This means that, even for galaxies closer than 5 Mpc, we need 
enough data points closer than 1\sec\ from the nucleus in order to observe a Keplerian rise of the velocity dispersion towards the nucleus ($\sigma\sim r^{-1/2}$), expected if it hosts a SMBH, limiting the feasibility of such studies to very nearby galaxies. There are very few close enough AGN's to be studied this way; besides, the nuclear regions of AGN's are very dusty, what means that optical observations cannot reach regions close enough to the SMBH; infrared observations should be used for this kind of study. As infrared instruments improve their performance, we should begin to see this kind of studies. Two additional difficulties with this method are: possible stellar velocity anisotropies in the nuclear region, which can mimic a Keplerian rise for $\sigma$, and projection effects (Kormendy \& Richstone 1995).

\subsection{Accretion Disks}

The strongest evidences for accretion disks have been 
obtained from gaseous kinematics. As the rotation of the accretion disk
is due to the SMBH inside, the evidences discussed below can
also be considered evidences for the nuclear SMBH, more so if they
are resolved and allow the determination of a lower limit for the]
nuclear mass density. 
Wilson (1996) and Ho (1998) present recent reviews about the gaseous kinematics around SMBH. A number of results build a consistent scenario:

\noindent
{\it 1) Observation of the Fe K$\alpha$ emission line profile in X-rays, at 6.4 keV}: about 20 Seyfert 1 galaxies observed with ASCA show broad Fe K$\alpha$
emission, with mean FWHM$\sim$50,000 km s$^{-1}$, with a strong red
wing indicative of gravitational redshifts close to a central black hole
(Nandra 1997).
Such results first appeared in the work of Tanaka et al. (1995), 
who found a double peaked  K$\alpha$ emission with width $\sim$ 100,000 km s$^{-1}$ in the nucleus of MCG-6-30-15, successfully modeled by emission from the inner region of an accretion disk (from 3 to 10 gravitational radii $R_G=(GM/c^2)^{1/2}$). Such observations are considered to be indirect evidence for the presence of a nuclear SMBH, as, although they are successfully modeled by gas orbiting a SMBH, they are not spatially resolved.

\noindent
{\it 2) Observation of double-peaked HI recombination lines}: $\sim$ 20 AGN are known to have such profiles (Eracleous \& Halpern 1994), with widths of $\approx$ 10,000 km s$^{-1}$. These profiles are successfully modeled by emission from the outer parts of an accretion disk (from 10$^2$ to 10$^4$ R$_G$) (Eracleous 1998); again, the observations are not resolved. 
An interesting case is that of of NGC\,1097.
This galaxy was known to present a LINER nucleus until Nov. 1991, when a double-peaked H$\alpha$ profile was first observed
(Storchi-Bergmann, Baldwin \& Wilson 1993). Since then we have followed
its variations, illustrated in Fig. 1 (Storchi-Bergmann et al. 1997). The clear double-peaked profile of the Balmer lines can only be observed when the stellar population is subtracted from the nuclear spectrum. Figure 2 illustrates this in a nuclear spectrum obtained on January 1998 with the Keck telescope by
A. V. Filippenko. One peculiar characteristic of the double-peaked profile in NGC\,1097 is that in some epochs the red peak is stronger than the blue peak, and, in a relativistic disk model, this can only happen if the disk is elliptical (Storchi-Bergmann et al. 1995).

\begin{figure}
\vspace*{0.25cm}
\begin{center}
\epsfig{file=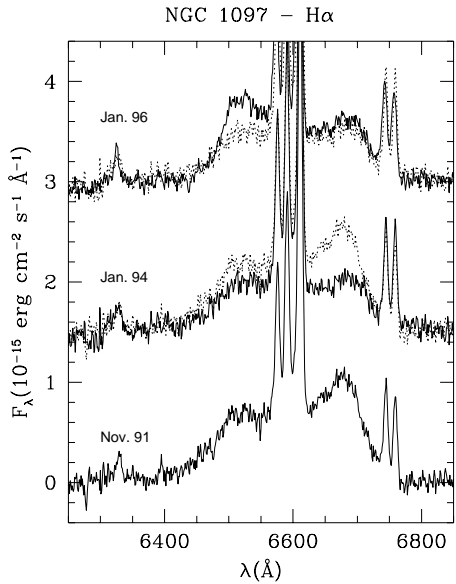, height=8.4cm}
\caption{NGC\,1097 double-peaked H$\alpha$ profiles in three epochs. Dashed lines show the profile from the previous epoch, for comparison (from Storchi-
Bergmann et al. 1997).}
\label{fig1}
\end{center}
\end{figure}

\begin{figure}
\vspace*{0.25cm}
\begin{center}
\epsfig{file=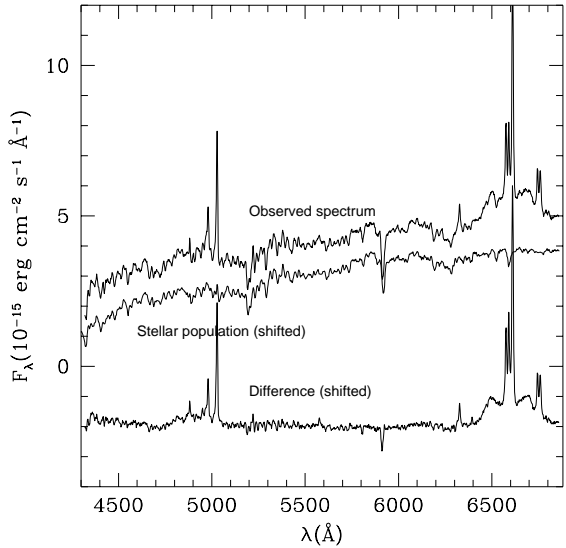, height=8.4cm}
\caption{Unpublished nuclear spectrum of NGC\,1097 obtained by Alex Filippenko
with the Keck telescope on Jan. 1998 (top). Also shown are 
the adopted stellar population template (middle) and the difference 
between the two (bottom). The two last spectra have been shifted for 
clarity.}
\label{fig2}
\end{center}
\end{figure}

\noindent
{\it 3) The H$_2$O Megamaser observed in NGC\,4258}: 
Miyoshi et al.(1995) have observed this galaxy in the 1.3cm maser-emission line of H$_2$O with the Very Long Baseline Array, achieving an angular resolution better than 0.5 miliarcseconds. The observations have resolved individual clouds in Keplerian rotation around the nucleus and located between  0.13 and 0.26 pc, corresponding approximately to 10$^5$ R$_G$, with maximum velocities at the inner radius reaching $\sim$ 1000 km s$^{-1}$. The inferred mass within this radius is $3.6\times 10^6 M_\odot$, too large for a long-lived star cluster with a radius smaller than 0.13 pc. This ring can either be identified with the
outer parts of the accretion disk or with the molecular torus.
Water vapor megamasers have been found in about 20 AGN's with a detection rate
of $\sim$6\%. This low detection rate is possibly due to anisotropy of the
maser emission or to a necessary combination of physical properties
(Wilson 1998 and references therein). From the 20, only a 
handful have high resolution
VLBI maps, and from those, in only half (4 cases) the emission seems to come from an edge-on rotating disc. But it was not possible to observe Keplerian rotation beyond the turnover in any other case besides NGC\,4258,
where the disc seems to be in a particularly favorable orientation 
(nearly edge on).

\noindent
{\it 4) HST nuclear disks images}: Optical images with the Hubble Space Telescope have revealed nuclear disks of gas and dust in scales of $\sim$100--300 pc in NGC\,4251, M\,87, NGC\,5322, NGC\,3415 AND M\,81 (Ho 1998 and references therein). The kinematics show keplerian motions with maximum velocities under 1000 km s$^{-1}$ (e.g. Machetto et al. 1997), and derived masses for the nuclear SMBH in the range 10$^8$--10$^9$ M$_\odot$. Although the derived nuclear densities are not huge, the corresponding mass-to-light ratios are usually very large, favoring the presence of a MDO in the nucleus. New HST observations with NICMOS at the emission line Pa$\alpha$ 
have also revealed an apparent 20 pc disk at the nucleus of Cen A 
(Schreier et al. 1998). This disk is, nevertheless, not oriented
perpendicular to the radio jet, possibly due to warping of its outer parts.

\section{SMBH in LINER's and normal galaxies}

According to Rees (1998) the most convincing case for a supermassive black hole using stellar kinematics, is the center of our own galaxy. Accumulating three-dimensional information on the movement of $\sim$ 200 individual stars in the galactic center in the IR, during 10 years, Eckart \& Genzel (1997) and Genzel et al. (1997) have obtained a Keplerian rise of the stellar velocities down to 1 light-week of Sagitarius A, deriving a mass for the central mass of $2.6\times 10^6$ M$_\odot$ and a corresponding density  larger than 
$2\times 10^{12}$ M$_\odot$ pc$^{-3}$.

LINER's can be considered intermediate cases between AGN's and normal galaxies. They were first identified as a class by Heckman (1980), for presenting a low ionization spectrum, and subsequent studies by Keel (1983) and Stauffer (1982) have shown that they are much more common than Seyfert galaxies in the nearby Universe. In most cases, they present little and faint emission. More recently, from a survey of nearby $\sim$ 500 galaxies, Ho et al. (1997) and Ho (1998) have concluded that $\sim$40\% are AGN's, 75\% of which are LINER's. From the LINER's, 20\% present broad H$\alpha$, approximately the same proportion observed in more luminous AGN's. These two latter results seem to confirm previous conclusions by Halpern and Steiner (1983) and Ferland \& Netzer (1983) who, using photoionization models to reproduce the emission lines of LINERþs, have concluded that most LINER's are AGN's  with a low ionization parameter.

Among the galaxies considered by Kormendy \& Richstone (1995) the best candidates for harboring a nuclear SMBH from stellar kinematics are normal galaxies or LINER's: NGC\,3115, NGC\,4486B, NGC\,4594, NGC\,3377, NGC\,3379, M\,31 and M\,32.
Candidates from more recent work with HST are M\,81, NGC\,3379, NGC\,4342 (Ho 1998 and references therein).

These results favor the idea that most present day luminous galaxies harbor a nuclear SMBH. This idea comes from the results of quasar counts as a function of redshift, which show that their number peak at z$\sim$ 2 (Rees, 1988). In terms of cosmic time, this result implies that there has been a ``quasar era'', with a relatively short duration, which occurred when the Universe was about 3-4 Gyrs old. The quasars from that era are the galaxies of today and, if every quasar harbored a SMBH,  we should expect that most luminous present day galaxies have a quiescent BH in their nuclei. The nucleus could become active by eventual accretion events occurring when a cloud of gas or a star passes close enough to the central hole to be captured (Rees 1988). This was the hypothesis put forth by 
Renzini et al (1995) to  explain the flare observed in ultraviolet images obtained with HST of the elliptical galaxy NGC\,4552, detected when the image was compared with another one obtained in a previous epoch. 

The proposed scenario for LINER's is that the nuclear SMBH has a low supply of gas, and thus the LINER's have low luminosities. Eventual accretion events could also occur in LINER's. This may have ocurred in the LINER nucleus of NGC\,1097 (Storchi-Bergmann et al. 1995): the transient double-peaked profile is produced in an elliptical accretion ring formed from the debris of the disruption of a star passing close to a nuclear SMBH with mass $\sim$ 10$^6$ M$_\odot$. Other double-peaked line appearances were also observed in M81 (Bower et al. 1996) and Pictor A (Halpern and Eracleous 1994) and could also be signatures of accretion events (Storchi-Bergmann et al. 1998).

\section{Feeding of the nuclear SMBH - relation to star formation}

In the scenario we are discussing here, the activity level is related to the accretion rate: the more luminous AGN's are the ones with larger accretion rates. LINER's are the AGN's with the lowest accretion rates. 

As discussed above, the accretion may occur via sporadic events, as is probably the case for NGC\,1097, NGC\,4552 and M\,81, or in a more steady way, if there is a large supply of gas. The latter case could occur, for example, following a massive star formation event close to the nucleus, when there would be plenty of mass released by the stellar formation and evolution processes.

A connection between star formation and nuclear activity has been proposed by a number of authors (e.g. Perry \& Dyson 1992). 
Terlevich, Diaz \& Terlevich (1990) have argued that the CaII triplet at $\sim\lambda$8500\AA\ in AGN's indicated the presence of young stars. 
Terlevich (1992) and Terlevich et al. (1992) have proposed what have since
been known as the Starburst Model for AGN. A recent review about this
topic has been presented by Cid Fernandes (1997). As an observational 
evidence for the presence of starburst in active galaxies,  
Cid-Fernandes \& Terlevich (1992, 1995) proposed that the second featureless continuum observed in Seyfert 2 galaxies (the FC2; Tran 1995) 
was due to young stars.
Recent studies using IUE and HST spectra and images have confirmed the presence of active star formation around the nuclei of a handfull of Seyfert 2 galaxies: Mrk\,477 (Heckman et al. 1997),
NGC\,7130, NGC\,5135, IC\,3639 (Gonzalez-Delgado et al. 1998). Storchi-Bergmann et al. (1998) have found another case --Mrk\,1210, discussed in more detail by Cid Fernandes et al. (1999).   

Yet another case seems to be that of the Seyfert 2 galaxy NGC\,7582,
which was found by Aretxaga (1999) and collaborators to present
transient broad profiles in the permitted nuclear emission 
lines. This galaxy has been known to present both Seyfert
and starburst characteristics (e. g. Kinney et al. 1993;
Cid Fernandes et al. 1998),
and the most probable interpretation for the origin of the broad lines
seems to be a supernova which have exploded close to the nucleus. This result is a nice confirmation of the viability of attributing at least some of the cases of rapid broad line variability (in particular, transient events) observed in Seyfert galaxies to supernova explosions (Terlevich et al. 1992).

 The star formation episodes in all the above
cases have ages between 3 and 6$\times 10^6$ yrs.
Signatures of the presence of young stars have been found also in LINER's. Maoz et al. (1998) found that at least half of the UV bright LINER's observed with HST ($\sim$ 25\% of the LINER's observed with HST are UV-bright) show  spectral fatures of young massive stars. Is this percentage significative? 
Does it differ from that observed in normal galaxies?

In order to investigate this issue for Seyfert galaxies, Schmitt, Storchi-Bergmann \& Cid Fernandes (1998; see also Cid Fernandes in these Proceedings) have analysed the nuclear spectra of 20 Seyfert 2 galaxies and 4 radio galaxies in comparison with that of an elliptical template through spectral synthesis. This comparison was motivated by the assumptions of previous works that the nuclear stellar population of AGN's can be reproduced by that of an elliptical galaxy. Figure 3 illustrates the synthesis results. It was concluded that in about 30\% of the sample there is larger contribution of stars younger than 10 Myrs than in the elliptical galaxy template. These ages correspond to cases like the individual Seyfert and LINER's discussed above. But the most significative result was that in 80\% of the sample, there was a larger contribution of stars of 100 Myrs than in the elliptical template. 

Storchi-Bergmann, Schmitt \& Cid Fernandes (1999) have compared the above statistics with that obtained for synthesis of integrated spectra of normal galaxies (Bica 1988) of the same Hubble types as the Seyferts (early type spirals). It was concluded that, for the normal galaxies, $\sim$20\% show larger contributions of stars younger than 10 Myrs than an elliptical template, a percentage only slightly smaller than that for the Seyferts, and similar to that in LINER's. This result suggest that the presence of stars younger than 10 Myrs in the nuclei of Seyfert 2 and LINER's is not necessarily linked to the nuclear activity, as non-active galaxies present similar statistical contributions of  such stars. On the other hand, only $\sim$20\% of the normal galaxies show contributions of 100 Myr stars in excess to that observed in ellipticals, as compared with $\sim$80\% for the Seyferts. This result suggests that star formation is linked to the nuclear activity through a 100 Myr timescale.
A more extensive discussion about this point can be found in the contribution
of Cid Fernandes et al.  to these proceedings.

\begin{figure}
\vspace*{0.25cm}
\begin{center}
\epsfig{file=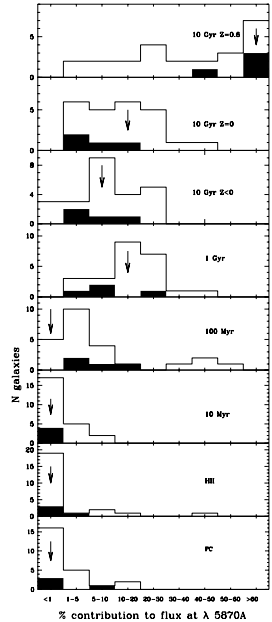, height=8.4cm}
\caption{From Schmitt et al. (1999): histogram of the different age and metallicity components contributions to the nuclear spectrum of 20 Seyfert 2 and 4 radio-galaxies, normalized to the light at $\lambda$5870\AA. Each panel
correspond to a different age, and the three upper ones separate also
the metallicity in three different bins. The filled histograms represent
the radio galaxies, while the open histogram represent the Seyferts plus
radio-galaxies.}
\label{fig:example}
\end{center}
\end{figure}

\section{Summary and concluding remarks}

I have reviewed the current evidences for the presence of a SMBH and surrounding accretion disk and torus in the nuclei of nearby AGN's, LINER's and normal galaxies. 

The strongest evidence for the presence of a SMBH using stellar kinematics comes from measurements of individual star speeds in the  Milky Way center, leading to the huge mass density of $\sim$10$^{12}$ M$_\odot$ pc$^{-1}$, which cannot be attained by a stable star cluster. The big advantage of the
galactic center is its proximity, which allows spatial resolution down to
a week light. A handfull of other candidates from (integrated) stellar kinematics include only one ``classical'' AGN (M87), the others being LINER's or normal galaxies. The main problem with the
nearby AGN is the large amount of dust, which precludes optical studies.
Future observations using infrared spectroscopy should increase the number of active galaxies in which these studies can be performed.

From gaseous kinematics, the strongest evidence of a nuclear 
SMBH comes from the masing disk in NGC\,4258, from measurements of 
resolved individual clouds in Keplerian motion down to 0.13 pc from the nucleus, also giving a mass density of $\sim$10$^{12}$ M$_\odot$ pc$^{-1}$.
This disk can  be identified with the outer parts of the accretion disk or
with the molecular torus of the standard model. 

Double-peaked emission-line profiles have been observed for the Fe 6.4 keV in X-rays, and for Balmer lines in the optical, and are thougth to be originated in the accretion disk. As these observations are not resolved, they are considered indirect evidence for the accretion disks. The widths of the profiles indicate that the Fe emission comes from only a few R$_G$ from the nucleus, while the widths of the optical Balmer lines indicate that they come from $\sim$10$^3$ R$_G$ from the nucleus (these radii are too small to be resolved). 

The fact that candidates for nuclear SMBH have been found in LINER's and normal galaxies, support the idea that there are SMBH in the nuclei of most luminous galaxies in quiescent state. The AGN activity would then correspond to a cycle in the life of the galaxy in which the central hole is accreting mass. The larger the accretion rate, more luminous is the AGN. The LINER activity corresponds to a low supply of gas. 

The feeding of the nuclear SMBH apparently occurs via (1) individual episodes, giving rise to flares or sudden appearance of double-peaked profiles; (2) more steady accretion, probably related to episodes of star formation. Recent results suggest a causal relation with a timescale or 100 Myr between a circumnuclear
starburst and the onset of nuclear activity. This maybe the long-sought link
between the starburst and nuclear activity.

\end{document}